\documentclass[12pt]{article}
\usepackage[pctex32]{graphics}
\begin{document}
\vspace{2cm}
\begin{center}
~\\
~\\
{\bf  \Large Thermal Instability of Giant Graviton \\
in Matrix Model on PP-wave Background }
\vspace{1cm}

                      Wung-Hong Huang\\
                       Department of Physics\\
                       National Cheng Kung University\\
                       Tainan,70101,Taiwan\\

\end{center}
\vspace{2cm}
\begin{center}{\bf  \Large ABSTRACT } \end{center}

The thermal instability of the giant graviton is investigated within the BMN matrix model.   We calculate the one-loop thermal correction of the quantum fluctuation around the trivial vacuum and giant graviton respectively.   From the exact formula of the free energy we see that at low temperature the giant graviton is unstable and will dissolve into vacuum fluctuation.   However, at sufficient high temperature the trivial vacuum fluctuation will condense to form the giant graviton configuration.   The transition temperature of the giant graviton is determined in our calculation.

\vspace{3cm}
\begin{flushleft}
E-mail:  whhwung@mail.ncku.edu.tw
\end{flushleft}


\newpage
\section{Introduction}
The matrix model on the eleven-dimensional maximally supersymmetric 
pp-wave background has been proposed by Berenstein, Maldacena and Nastase [1,2], which is often referred as the BMN matrix model.  The model provides an interesting example where the string/gauge correspondence can be studied beyond the supergravity approximation [3,4].   Besides the terms in the BFFS matrix model [5] the BMN matrix model also contains mass terms and Myers terms [6].   The extra terms can remove completely the moduli space and leave only the discrete set of solutions.   Due to the Myers mechanics the RR-flux (i.e., the extra terms)  in the BMN model can expand the brane into its spherical form, i.e., the fuzzy spheres [1,2,6].   One of the sphere is the non-supersymmetric fuzzy sphere while the another one is the supersymmetry (1/2 BPS) fuzzy sphere which is called as giant graviton [7].   Unlike the similar situation studied in Myers [6] the all giant graviton solutions have zero energy and is independent of the matrix dimension and representation of SU(2).

   In [8] Sugiyama and Yoshida had used the background field method to calculate the one-loop quantum correction of the BMN matrix model.   They found that there does not induce any correction around the supersymmetric fuzzy sphere and thus showed the quantum stability of the giant graviton.   In this brief report we will use their formulations to evaluate the one-loop thermal correction of the BMN matrix model.  We evaluate the free energy of  the quadratic fluctuations about the trivial vacuum and giant graviton respectively.   We see that while the trivial vacuum and giant graviton solution both have zero energy and does not induce any correction at zero temperature [8] they have different free energy at finite temperature.   Our result shows that giant graviton has higher free energy than that of the trivial vacuum fluctuation at low temperature.
Thus at low temperature the giant graviton will be unstable and dissolve into the trivial vacuum fluctuation.  However,  at sufficient high temperature the giant graviton will have less free energy than that of the trivial vacuum fluctuation.   This means that the trivial vacuum fluctuation will condense into the giant gravitons which now become the stable configurations.  We also determine the transition temperature of the giant graviton in our calculation.

   In section II  we briefly review the formulations and results of Sugiyama and Yoshida [8].  In section III we adopt these formulations to evaluate the quantum correction at finite temperature system and drown the conclusions in the last section.

\section{BMN Matrix Model and Quantum Fluctuation}
The matrix theory on the eleven-dimensional maximally supersymmetric pp-wave background is described by the BMN matrix model.    The action is given $ S=S_{flat} + S_{\mu}$ where
$$S_{flat}=  \int dt ~Tr \Biggl[{1\over 2}D_0 X^r D_0 X^r + \frac{1}{4}[X^r, X^s]^2 + i \Psi^T D_0 \Psi - \Psi^T  \gamma_r [\Psi, X^r] \Biggr], ~~ r,s=1,2,...,9, \eqno{(2.1)}$$
$$S_{\mu} = \int dt ~Tr \left [ - \frac{1}{2} \left(\frac{\mu}{3}\right)^2 X_ I^2 - \frac{1}{2}\left( \frac{\mu}{6}\right)^2 X_{I'}^2 - i\frac{\mu}{3}\epsilon_{ IJK}X^{I}X^{J}X^{ K} - i\frac{\mu}{4}\Psi^{T}\gamma_{123}\Psi \right],~~~ I=1,2,3,~~~$$
$$ \hspace{12cm} I' = 4,5,...,9, \eqno{(2.2)}$$
where the covariant derivative is defined by  $ D_0 X^r \equiv \partial_t X^r - i[A, X^r]$ and we have rescaled the gauge field $A$, parameters $t$ and $\mu$ as $  t \rightarrow \frac{1}{R} t , ~~~ A \rightarrow R A , ~~~ \mu \rightarrow R \mu $ in which $R$ is the radius of circle compactification along $x^-$ [1].

   To proceed, we first decompose the $X^r$ and $\Psi$  into the backgrounds $B^I (= \alpha J^I), F$  and fluctuations $Y^r, \psi$ as 
$$ X^r = B^r + Y^r ,~~~  \Psi = F + \psi ,  \eqno{(2.3)}$$
where we take $F=0$ since the fermionic background is not considered in this paper.   When $\alpha = {\mu\over 3}$ the background is the giant graviton while the case of $\alpha =0$ is the trivial vacuum solution.  Note that the giant graviton is the solution of  (2.1) and has zero energy like as the trivial solution [1,2].  Next, we adopt the background field gauge and the gauge-fixing terms and Faddeev-Popov ghost terms are
$$  S_{GF+FP} = - \frac{1}{2}\int dt~ Tr \left[  (D_0^{bg}A)^2 + i \bar{C}\partial_t D_0 C + \bar{C}[B^r,[X^r,\,C]] \right ], \eqno{(2.4)}$$
where $D_0^{bg}A$ is defined by $ D_0^{bg}A= \partial_t A  + i[B^r,X^r]$.
The advantage of this gauge choice is that the second order actions with respect to the fluctuations are simplified [8]. 

  For the case of  $2\times 2$ matrices we can expand the fluctuations and gauge fields as 
$$ Y^r = \frac{1}{\sqrt{2}}Y_0^{r} {\bf 1}_2 + \sqrt{2}Y_1^{r}J^1 + \sqrt{2}Y_2^{r}J^2  + \sqrt{2}Y_3^{r}J^3 , $$
$$ \psi =  \frac{1}{\sqrt{2}}\psi_0 {\bf 1}_2 + \sqrt{2}\psi_1 J^1 + \sqrt{2}\psi_2 J^2  + \sqrt{2}\psi_3J^3, $$
$$ A = \frac{1}{\sqrt{2}}A_0{\bf 1}_2 + \sqrt{2}A_1J^1 + \sqrt{2}A_2J^2 
+ \sqrt{2}A_3J^3, \eqno{(2.5)}$$ 
where $J^I \equiv \sigma^I /2,  (I = 1, 2, 3) $  and $\sigma^{I}$'s are Pauli matrices.  Substituting these terms into (2.1) and (2.2) then after the diagonalization procedure the quadrature fluctuations about the giant graviton have the following contents of the fields and spectra:   
\\
\\
\begin{tabular}{|c|c|c|c|c|}
\hline
fields&bosonic fluctuations $Y$ &gauge fields $A$&ghost fields $C$&fermions $\psi$\\
\hline
degeneracy&4~~~~~3~~~~5~~~ 6~~~18 &1~~~~~~3 &1~~~~~~3 &16~~~32 ~~~16\\
\hline
mass&${\mu \over 3}$~ ${{\sqrt 2}\mu \over 3}$~ ${2 \mu \over 3}$~~ ${\mu \over 6}$~~~${\mu \over 2}$ &~~~0~~~~${{\sqrt 2}\mu \over 3}$ &~~0~~~~${{\sqrt 2}\mu \over 3}$&$~{\mu \over 4}$~~~~${7\mu \over 12}$~~~${-5\mu \over 12}$\\
\hline
\end {tabular}
\\

~~~ Table 1.   The spectrum of fluctuation in giant graviton.\\
After integrating the quadratic fluctuations the one-loop effective action $W$  has four contributions.   The contribution of the fluctuations $Y$, gauge field $A$, the ghost $C$ $(\bar{C})$ and fermion parts $F$ are respectively given by

$$ W_{Y} =  - \ln \Biggl[\rm  Det\left[ - \partial_{\tau}^2 + \frac{\mu^2}{9} \right]^{-4/2} {\rm Det} \Biggl[- \partial_{\tau}^2 + \frac{2}{9}\mu^2 \Biggr]^{-3/2} {\rm Det} \Biggl[- \partial_{\tau}^2 + \frac{4}{9}\mu^2 \Biggr]^{-5/2} $$
$$~~~~~~~~~~~~~~~~~~~~~~~\times{\rm Det} \left[- \partial_{\tau}^2 + \frac{\mu^2}{36} \right]^{-6/2} {\rm Det} \left[- \partial_{\tau}^2 + \frac{\mu^2}{4} \right]^{-18/2} \Biggr].  \eqno{(2.6)}$$

$$ W_{A} = - \ln {\rm Det} \Biggl[- \partial_{\tau}^2 + \frac{2}{9}\mu^2 \Biggr]^{-3/2}, ~~~~~ W_{\rm gh} = - \ln {\rm Det} \Biggl[- \partial_ {\tau}^2 + \frac{2}{9}\mu^2 \Biggr]^{3}.\eqno{(2.7)}$$

$$ W_{\rm F} = - \ln \Biggl[ {\rm Pf} \Biggl[ i\partial_{\tau} + \frac{\mu}{4}\gamma_{123} \Biggr] {\rm Pf} \Biggl[ i\partial_{\tau} + \frac{7}{12}\mu\gamma_{123} \Biggr]^2 {\rm Pf} \Biggl[ i\partial_{\tau} - \frac{5}{12}\mu\gamma_{123} \Biggr] \Biggr], \eqno{(2.8)} $$
\\
where $[{\rm Pf}(B)]^2 = {\rm Det}(B)$.    Finally, using the property derived in [8]
$$  \ln{\rm Det}[-\partial_{\tau}^2 + M^2] = L \int \! \frac{dk_0}{2\pi}  \ln (k_0^2 + M^2) = L(|M| + E_{\infty}), \eqno{(2.9)}$$
$$\ln {\rm Pf}\left[ i\partial_{\tau} + M\gamma_{123} \right] = 4L \int \! \frac{dk_0}{2\pi}  \ln (k_0^2 + M^2) =  4 L (|M| + E_{\infty}), \eqno{(2.10)}$$
where $L$ is the length of the temporal direction and 
$$ E_{\infty} = \int {dk\over 2\pi}~ ln k^2,  \eqno{(2.11)}$$
the net one-loop contribution is exactly canceled  as expected from the supersymmetry.    This is the result of [8].

   To proceed we need to know the quadrature fluctuations about the trivial vacuum.    Following the above procedure we  have the following contents of the fields and spectra:   
\\
\\
\begin{tabular}{|c|c|c|c|c|}
\hline
fields&bosonic fluctuations $Y$ &gauge fields $A$&ghost fields $C$&fermions $\psi$\\
\hline
degeneracy&12~~~~~~24 &4 &4 &64\\
\hline
mass&${\mu \over 3}$~~~~~ ${\mu \over 6}$&0&0&${\mu \over 4}$\\
\hline
\end {tabular}\\
\\ 
Table 2.   The spectrum of fluctuation in trivial vacuum.
\\
From the above spectra the net one-loop corrections in the trivial vacuum can also evaluated in the same way.    It is easy to see that they are exactly canceled as expected from the supersymmetry. 

\section{One-Loop Thermal Correction and Free Energy}
In finite temperature $T (= 1/\beta)$ the integration of $k_0$ in (2.9) and (2.10) shall be replaced by a summation over $2n\pi/\beta $.  The summation can be performed by the following relation.

   For the case of boson $n$ is the nature integral (N.I.) and we have the relation 
$$\sum_{n\in N.I.} \ln \left[(2\pi n)^2+(\beta M)^2\right] = \sum_{n\in  N.I.}  \int_1^{(\beta M)^2}{ ds^2 \over (2\pi n)^2+ s^2} + \sum_{n\in  N.I.}  \ln \left[1+ (2\pi n)^2\right]$$
$$= \int_1^{(\beta M)^2} {ds^2 \over 2s} \Biggl[1+ {2 \over e^s -1}\Biggr] + \sum_{n\in  N.I.}  \ln \left[1+ (2\pi n)^2\right] \hspace{3.5cm}$$
$$= (\beta M-1) + 2 \left[\ln (1- e^{-\beta M}) - \ln (1- e^{-1})\right] + \sum_{n\in  N.I.}  \ln \left[1+ (2\pi n)^2\right]\eqno{(3.1)}$$
In the same way, for the case of fermion  $n$ is the half integral (H.I.) and we have the relation 
$$\sum_{n\in H.I.} \ln \left[(2\pi n)^2+(\beta M)^2\right] = \sum_{n\in  H.I.}  \int_1^{(\beta M)^2}{ ds^2 \over (2\pi n)^2+ s^2} + \sum_{n\in  H.I.}  \ln \left[1+ (2\pi n)^2\right]$$
$$= \int_1^{(\beta M)^2} {ds^2 \over 2s} \Biggl[1- {2 \over e^s +1}\Biggr] + \sum_{n\in  H.I.}  \ln \left[1+ (2\pi n)^2\right]\hspace{3.5cm}$$
$$= (\beta M-1) + 2 \left[\ln (1+ e^{-\beta M}) -  \ln (1+ e^{-1})\right] + \sum_{n\in  H.I.}  \ln \left[1+ (2\pi n)^2\right].\eqno{(3.2)}$$
Then using the spectrum in the table 1 and 2 we can easily calculate the partition function and thus the free energy.   The exact formula of the partition function is

$$\ln Z = \sum_{i \in Y, A}({-N_i\over2})\Biggl[ (\beta m_i-1) + 2 \left[\ln (1- e^{-\beta m_i}) - \ln (1- e^{-1})\right] + \sum_{n\in  N.I.}  \ln \left[1+ (2\pi n)^2\right]\Biggr]$$
$$ + \sum_{i \in C,\psi}\left({N_i \delta_{i,C} }+{N_i \delta_{i,\psi} \over4} \right) \Biggl[ (\beta m_i-1) + 2 \left[\ln (1+e^{-\beta m_i}) - \ln (1+ e^{-1})\right] + \sum_{n\in  H.I.}  \ln \left[1+ (2\pi n)^2\right]\Biggr],\eqno{(3.3)}$$
in which $N_i$ and $m_i$ are the degeneracy and mass of the fluctuation field type $i$.   After the numerical evaluation we have seen that the free energy of fluctuation from the trivial vacuum is lower than that from the giant gravity system at low temperature.  However when the temperature is larger than the critical temperature $T_c$ 
 $$ T_c ~~\approx ~~385 \mu,  \eqno{(3.4)}$$
then the free energy of giant graviton becomes less than that of the trivial vacuum. 
    The analytic relations can be obtained in the high-temperature and low-temperature cases.  For example, in the high temperature limit we have a simple relation 
$$  F_{giant~graviton}  -  F_{trival ~vacuum} \approx  ~ -3 ~ T \ln (T/ \mu), ~~~~~T \gg \mu . \eqno{(3.5)}$$
While in the low temperature limit we have
$$  F_{giant~graviton}  - F_{trival ~vacuum} \approx  ~ 18~ T ~e^{- \mu/6T}, ~~~~~T \ll \mu . \eqno{(3.6)}$$
We thus conclude that at high temperature the vacuum fluctuation will condense into the configuration of giant graviton.   On the other hand, at low temperature the giant graviton will become unstable and dissolve into vacuum fluctuation.   
\\
\hfil\scalebox{1}{\includegraphics{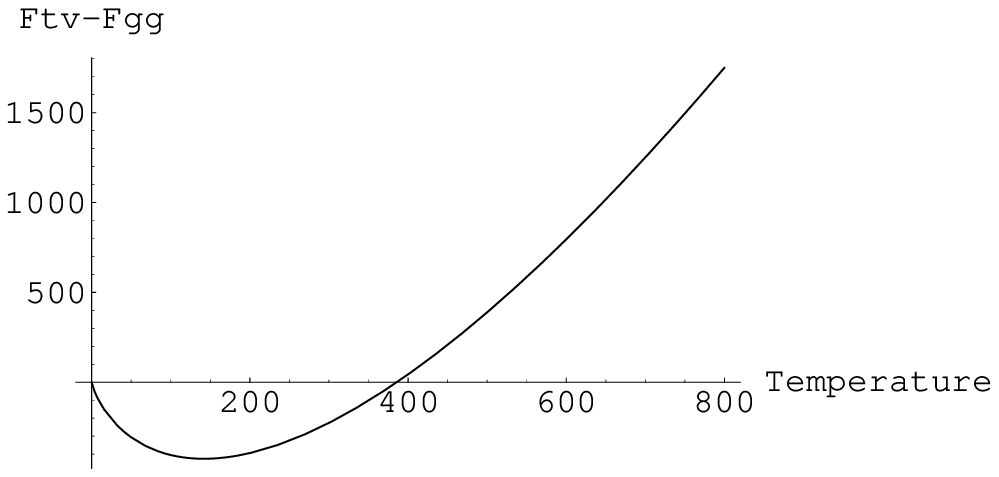}}\hfil\\
{\it ~~~Fig.1. Temperature dependence of  the  difference between the trival-vacuum free energy Ftv and giant-graviton free energy Fgg.  The scale is in the unit of $\mu$ = 1.}
\\
\section{Conclusion}
   In this paper we use the formulations of  Sugiyama and Yoshida [8] to evaluate the one-loop thermal correction in the BMN matrix model.  We evaluate the free energy of  the quadratic fluctuations about the trivial vacuum and giant graviton respectively.   We see that while the trivial vacuum and giant graviton solution both have zero energy and does not induce any correction at zero temperature [8]  the trivial vacuum will have less free energy than that in giant graviton during increasing the temperature.   Therefore at low temperature the giant graviton will be unstable and dissolve into the trivial vacuum fluctuation.  However,  at sufficient high temperature the giant graviton will have less free energy than that of the trivial vacuum fluctuation.   This means that the trivial vacuum fluctuation will condense into the giant gravitons which now become the stable configurations at high temperature.   The phenomena of a physical configurations which become stable at high temperature can also be found in other systems.  The simplest one may be that the topological vortex will condense at high temperature in the 2-dimensional XY statistical model [9]. 

    It is hoped that the property of phase transition of giant graviton we find may be relevant to the cosmological evolution in which the temperature is cooling down during the expansion of universe.

   Finally we want to mention that Bak [10] had dealt with the giant
graviton and its elliptic deformation with 8 SUSY. The computation of free
energy of these elliptic configurations and investigation of their thermal stability would be quite interesting.  It remains to be studied.

\newpage
{\bf  \Large REFERENCES}
\begin{enumerate}
\item  D.~Berenstein, J.~Maldacena and H.~Nastase, ``Strings in flat space and pp waves from  N= 4  Super Yang Mills'', JHEP  0204 (2002) 013,  hep-th/0202021.
\item K.~Dasgupta, M.M.~Sheikh-Jabbari and M.~Van Raamsdonk, 
``Matrix Perturbation Theory For M-Theory On a PP-Wave'', JHEP  0205 (2002) 056,  hep-th/0205185.  
\item  K.~D. Sadri and M.M.~Sheikh-Jabbari, ``The Plane-Wave/Super Yang-Mills Duality'', Rev. Mod. Phys. ,  hep-th/0310119.  
\item   J.~Maldacena, ``TASI 2003 Lectures on AdS/CFT'', hep-th/0309246.  
\item  T.~Banks, W.~Fischler, S.H.~Shenker and L.~Susskind, 
``M-theory As A Matrix Model: Conjecture'', Phys. Rev. , D55 (1997) 5112,  hep-th/9610043.
\item R. C.~Myers, ``Dielectric-Branes'',  JHEP  9912 (1999) 022,  hep-th/9910053. 
\item  J.~McGreevy, L.~Susskind and N. Toumbas ``Invasion of Giant Gravitons from Anti de Sitter Space'', JHEP  0006 (2000) 008,  hep-th/0003075;  M. T. Grisaru, R. C.~Myers and O. Tafjord, ``SUSY and Goliath'', JHEP  0008 (2000) 040,  hep-th/0008015. 
\item  K.~Sugiyama and K.~Yoshida, ``Giant graviton and quantum stability in matrix model on PP- wave background,'' Phys. Rev.  D66 (2002) 085022, hep-th/0207190.
\item J. B. Kogut, ``An Introduction to Lattice Gauge Theory and Spin Systems'', Rev. Mod. Phys. 51 (1979) 659.
\item D. Bak, ``Supersymmetric Branes in the Matrix Model of PP Wave Background,'' Phys. Rev. D67 (2003) 045017, hep-th/0204033.
\end{enumerate}

\end{document}